# Cell shape identification using digital holographic microscopy

JOHAN ZAKRISSON,[1] STAFFAN SCHEDIN,[2] AND MAGNUS ANDERSSON[1,*]

[1]Department of Physics and [2]Department of Applied Physics and Electronics, Umeå University SE-901 87 Umeå, Sweden
*Corresponding author: magnus.andersson@umu.se

**We present a cost-effective, simple and fast digital holographic microscopy method based upon Rayleigh-Sommerfeld back propagation for identification of the geometrical shape of a cell. The method was tested using synthetic hologram images generated by ray-tracing software and from experimental images of semi-transparent spherical beads and living red blood cells. Our results show that by only using the real part of the back-reconstructed amplitude the proposed method can provide information of the geometrical shape of the object and at the same time accurately determine the axial position of the object under study. The proposed method can be used in flow chamber assays for pathophysiological studies where fast morphological changes of cells are studied in high numbers and at different heights.**

## 1. Introduction

Digital holography microscopy (DHM) is an evolving coherent imaging technology that is experiencing a tremendous growth and has potential applications in a wide range of areas including; cellular microscopy, manufacturing processes, medical imaging, biometry, and environmental research [1]. DHM is an efficient and easy-to-operate technique that allows obtaining from single recorded holograms, quantitative phase images of living cell dynamics with interferometric resolution i.e., with an axial sensitivity at tens of nm [2]. The advantage of holography compared with normal microscopic imaging is that the hologram contains all the information necessary to reconstruct the amplitude and phase of a propagating light field at different spatial positions, i.e., focusing of the object image at a well-defined plane [3]. The reconstructions are performed numerically by using fast 2D-Fourier transforms. The most common methods for reconstruction are the Fresnel transform method, the Huygens convolution method and the angular spectrum method [1, 4]. In addition, the reconstruction procedure has also been extended to include self-focusing for optimization of reconstructed images [5, 6].

Digital holography can thus be used for quantitative high-precision deformation measurements and measurements of refractive index distribution of semi-transparent macroscopic objects [7]. In addition, DHM has particularly been proven as a powerful technique in microbiology, where it can be used to assess parameters such as thickness, volume and refractive index of living as well as dead cells [8, 9]. This non-invasive and staining-less technique has therefore helped elucidate cell movement in both 2D and 3D, as well as shed light on cell morphology, both at a single cell level and for entire populations [10]. Understanding cell morphology is important from a fundamental point of view since the shape of a cell can change in response to several different stimuli, and since it also relies on intracellular mechanics [11]. In addition, the cell's physical interaction with its environment and the possibility to change its morphology using drugs for treatment of diseases add momentum to the development of non-invasive imaging techniques such as DHM [12].

In this work we demonstrate a fast, low computational intense method based upon DHM, which can be used for analyzing morphological changes of cells and for tracking their axial position. Using the real part of the back-reconstructed amplitude (containing parts of the intensity and phase), we study how this varies for spherical and elliptical shaped objects.

## 2. Experimental procedure and theory

We built our setup around an Olympus IX71 microscope, normally used for optical tweezers force spectroscopy and flow chamber assays, which was modified to illuminate samples according to an in-line holography setup using a collimated low-cost LED operating at 470 nm (M470L3-C1, Thorlabs) [13]. The LED was collimated using a free space collimator consisting of a microscope objective (Plan N 10x/0.25, Olympus) mounted on a fiber holder (MBT613, Thorlabs). The objective focused the light from the LED into a multi-mode (MM) fiber (QMMJ-3AF3AF-IRVIS-50/125-3-5, OZ Optics) and an output collimator (HPUFO-2, A3A-400/700-P-17-180-10AC, OZ Optics) was mounted just above the piezo-controlled microscope stage holding the sample (PI-P5613CD, Physik Instruments) [14]. The MM fiber allowed for frequency filtering of the light and for easy and stable arrangement of the illumination source. The stability of the setup was optimized to reduce drift and noise by measuring long time series and by using the Allan variance approach described in [15].

### 2.1 Sample preparation and measurement

Samples were prepared by mixing polystyrene beads with a diameter of 9.685 μm (4210A, Thermo Scientific) and filtered Milli-Q water to preferred concentrations. RBCs were prepared by first mixing one drop of human blood with 500 μl PBS pH 7.4. This suspension was centrifuged at 2000 rpm for 2 minutes and the supernatant was replaced with new PBS. This procedure was repeated three times to remove any influence of serum proteins.

A sample chamber was prepared by applying two pieces of double sticky tape, parallel to the long side of a 24x60 mm coverslip (no.1, Knittel Glass), such that a channel with a spacing of 5 mm was formed. On top of the tape a 20x20 mm coverslip (no.1, Knittel Glass) was placed forming a closed channel. This method is well proven and reliable and has been used extensively in bacterial force spectroscopy studies as well as for tethered particle tracking studies [16–18]. The solution of ~10 μl with either beads or RBCs was added at one of the openings of the channel and via capillary forces the channel was thus loaded. Then openings of the sample chamber were sealed by vacuum grease (DOW CORNING®) to avoid drying. Finally, the chamber was placed in the microscope stage and the beads/RBCs were incubated for 15 min to immobilize to the glass surface.

A sample was thereafter illuminated with the LED from the top and imaged using a 60x water immersion objective (UPlanSApo 60x, Olympus), as schematically shown in Fig. 1. The use of a water immersion objective reduces the amount of aberrations allowing for better quality of the hologram image. The position of the sample was controlled by the piezo-stage, which could be translated 100 μm along all axes. Finally, a 12-bit CCD-camera (DX 2 HC-VF, Kappa) with a SNR of 63 dB was used to capture the holograms at various distances behind the object. The camera was positioned in the x-y plane parallel to the hologram plane.

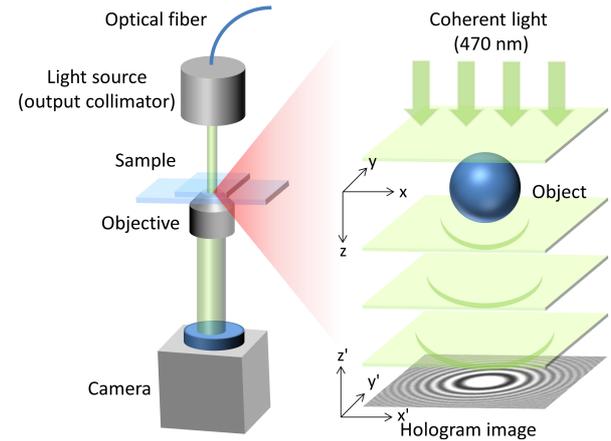

Figure 1. The experimental setup with light paths and optical components is shown to the left whereas a zoom-in of a spherical object, illuminated with a coherent light source, is shown to the right with a resultant hologram. We used an in-line holographic microscope approach with a blue LED as light source operating at 470 nm. The origins for the $x, y, z$ and $x´, y´, z´$ coordinate systems were positioned at the center of the object and at the center of the hologram plane, respectively. Thus, the back-reconstruction was done from the hologram image along the z´-direction towards the object.

### 2.2 Theory

The original in-line version of holography, referred to as Gabor holography, can shortly be explained as follows [19]. Consider a small semi-transparent particle illuminated by a plane wave. The particle will scatter a fraction of the incident light field and at a distance $r$ from the particle a modulation of the irradiance pattern, due to the interferences of the scattered and non-scattered fields, can be recorded on a detector. The irradiance, $I$, recorded on the detector (i.e., the hologram) is then given by,

$$I = |U_O + U_R|^2 = |U_O|^2 + |U_R|^2 + U_O U_R^* + U_O^* U_R, \qquad (1)$$

where $U_O$ and $U_R$ are the scattered and the non-scattered (reference) amplitude, respectively, and * denotes a complex conjugate. The scattered amplitude is normally much smaller than the amplitude of the reference, implying that the first term of Eq. (1) is negligible. Using the Rayleigh-Sommerfeld theory of light propagation it can be shown that the diffracted field, $U$, can be solved by convolution with the complex amplitude impulse response function, $h$, defined as:

$$h = \frac{z'e^{jkr}}{j\lambda r^2}, \qquad (2)$$

where $z'$ is the axial distance from the hologram plane (image plane) to the reconstruction plane, $k$ is the wave number, $j$ is the complex number, and the distance $r$ is given by $r = \sqrt{z'^2 + (x'-x)^2 + (y'-y)^2}$ where $x, y,$ and $x', y'$ are the lateral co-ordinates at the hologram and reconstruction planes, respectively [19]. The diffraction integral can thus be expressed as:

$$U(x',y';z') = \iint U(x,y) \cdot h(x'-x, y'-y; z') \, dx \, dy. \qquad (3)$$

This integral can be numerical evaluated by using the fact that the convolution of the product $I \cdot h$ is equal to the inverse Fourier transform of the product of the individual Fourier transforms of $I$ and $h$,

$$U(x',y';z') = \mathbf{F}^{-1}\left[\mathbf{F}\{U(x,y)\} \cdot \mathbf{F}\{h(x'-x, y'-y; z')\}\right]. \qquad (4)$$

The Fourier transform of the impulse response function, $\mathbf{F}\{h\}$, is referred to as the transfer function of free space propagation, and is in the angular spectrum approach given by,

$$\mathbf{F}\{h(x',y';z')\} = \exp\left[\frac{2\pi j z'}{\lambda}\sqrt{1 - (\lambda f_x)^2 - (\lambda f_y)^2}\right], \qquad (5)$$

where $f_x$ and $f_y$ are the spatial frequencies [19].

## 3. Results

### 3.1 Validation of the algorithm and axial position determination of spherical objects

We implemented the algorithm in Matlab, and for validation we generated well defined synthetic hologram images using Zemax simulation software. In the simulation, a spherical object with a diameter of 9.685 μm and with an index of refraction of 1.52 was illuminated with a coherent light source of wavelength 470 nm, thus similar to the experimental setup. The virtual light source used 4.04 billion individually photons that were traced in each simulation. Virtual detectors (treated as transparent) located at specified z positions behind the object counted the number of transmitted photons and their phase respectively, resulting in synthetic holograms. The distances were 30, 40, 50, 60, 70, 80, 90, and 100 μm behind the particle, thus representing a set of image planes assessed with an objective in a microscope. We denote the distance from the hologram plane (image) to the reconstructed plane to z´. An example of a synthetic hologram at a distance of 80 μm from the sphere (thus defined as z´= 0 and z = 80 μm) is shown in Fig. 2A.

Next, we performed a set of physical experiments where 9.685 μm polystyrene spheres were immobilized to coverslips and illuminated with the LED at 470 nm in our modified IX71 microscope. A hologram of a representative sphere recorded at 80 μm is shown in Fig. 2C. The cover slip was translated along the optical axis (i.e., positive z-direction) by moving the piezo-stage in discrete 10 μm steps (from 30 μm to 100 μm) and a hologram (image) was acquired at each step.

The simulated and experimental hologram images were back-reconstructed with the in-house Matlab algorithm, in which the theoretical description of light propagation was implemented as described above. The reconstructed field (given by Eq (4)) was evaluated at a large number of cross-sectional planes (x´-y´-planes) along the z´-axis (from 0 μm to 100 μm with a spacing of ∼Δz´ = 70 nm) in between each image. Along the optical axis, the intensity profile was thereafter extracted from the data and further analyzed. The result of such analysis is shown in Fig. 2B and 2D where the normalized intensity profile, reconstructed from each image from the simulated and the measured data, respectively, are presented as a function of distance from the hologram. In Fig. 2E the z´-position of the maximum value of the intensity profiles, representing the position of the geometrical focus, in Fig. 2B and D, are plotted versus the real distance. As can be seen, the linear fits to the data, red and green line with $R^2$ values of 0.99977 and 0.99996, respectively, indicates that the reconstruction algorithm can perfectly predict the position of both the simulated and experimental data. The top panel in the inset in Fig. 2E shows the Gouy phase shift of ∼π over the focus [20], and the lower panel shows the intensity along the optical axis (for the hologram at 80 μm). Thus, this allowed for accurate axial position quantification of a spherical object.

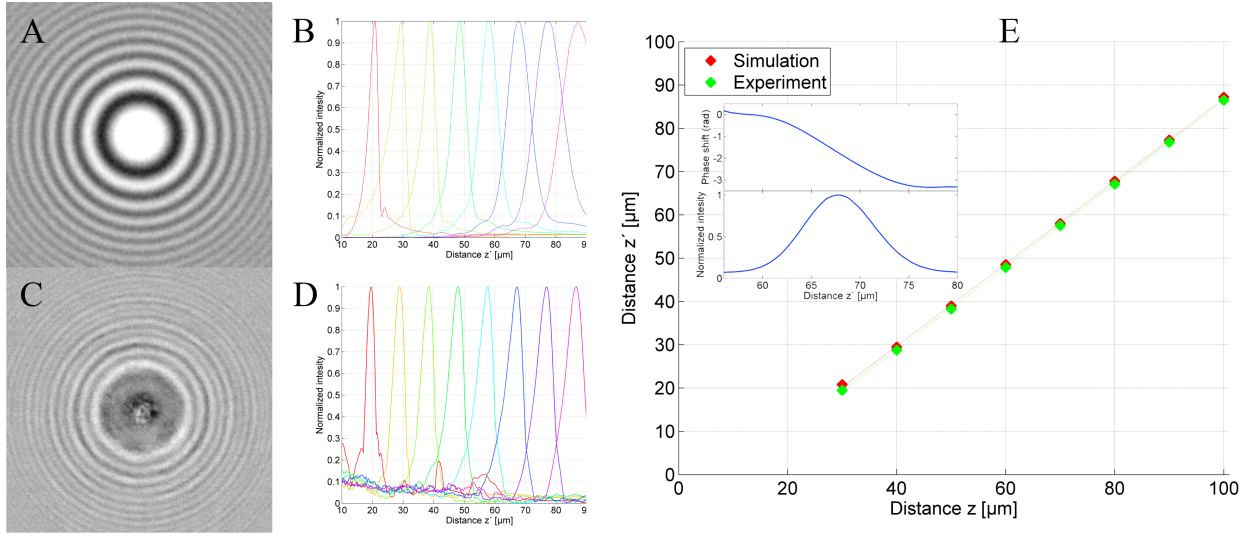

Figure 2. Hologram images of simulated and experimental data. Panel A) shows the hologram of a synthetic image acquired 80 μm behind the 9.685 μm spherical object. B) the reconstructed absolute intensities (eight positions) along the optical axis for the simulated sphere. C) image acquired 80 μm behind a 9.685 μm polystyrene bead. D) the reconstructed absolute intensities along the optical axis in the experiment. E) Piezo-position vs. reconstructed position of the focus. Red dots represent the reconstructed positions of the simulation and the red line is a linear fit to the data. Similarly, the green dots represent the experimental data with the corresponding linear fit. The inset shows the Gouy phase shift at the reconstructed focus for the image acquired at 80 μm behind the object in the simulation.

**3.2 Identifying morphological changes**

To scrutinize if it is possible to identify morphological changes using phase information and reconstructed data, we first generated synthetic holograms with objects of various degrees of ellipticity. A sphere of 7 μm radius, a large ellipse (7x5 μm radii) and small flat ellipse (7x3 μm radii) with an index of refraction of 1.50 were simulated in Zemax, with the same setup and illumination conditions as described above. The ellipses were positioned with its short radius along the optical axis. The conic constants were set to 0.96 and 4.44 whereas the radii were 8.17 μm and 4.9 μm, for the large and small ellipse, respectively, see ref. [21], for details. Again, synthetic holograms were generated at fixed 10 μm steps along the optical axis, from 30 to 100 μm behind the objects. The maxima of the back-reconstructed intensity profiles from the synthetic holograms from each of the three objects are presented in Fig. 3A. As can be seen, for each reconstructed set the data is linear with similar slope but shifted in the vertical direction. The lines represent linear fits with $R^2$ values of 0.99993, 0.99952, and 0.99038 for the sphere, large ellipse and small ellipse, respectively. This implied that a morphological change, i.e., a change from a spherical to an elliptical shape, cannot be identified by studying the reconstructed intensity profile only, since it is not possible to identify if the vertical shift is a result of a change of the objects shape or if it has moved in the z-direction. In addition, the phase shift around the focus is ~π radians, i.e., the Gouy phase anomaly, and will thereby not provide information about the shape. Therefore, we investigated if it was possible to study the real part of the amplitude Re($U$) around the geometrical focus of the back-reconstructed light to identify specific geometrical features for a given object.

We studied Re($U$) of the light field along the optical axis in a similar manner as found in refs. [22, 23]. Fig. 3B and 3C, show Re($U$) from the back-reconstruction of the holograms (acquired at positions 70 μm and 100 μm from the object) along the optical axis for the sphere (red curve), the large ellipse (green), and the small ellipse (blue). The position of the object is indicated by a grey vertical line. Clearly, the three curves exhibit distinct difference of Re($U$) around the geometrical focus, which is represented by * in Fig. 3B and 3C and located at distances ~11, ~18 and ~38 μm for the sphere, large ellipse, and small ellipse, respectively. The major peak found in the red curve (spherical object) changes into a lower and broader peak, and moves closer to the hologram plane, as the object becomes flatter. Also, a "dip" is located before the peak at ~20 μm for the sphere, whereas for the large ellipse (green curve) there is no apparent "dip" and for the small ellipse (blue curve) a "dip" is located after the peak (same location as for the sphere). This implied that by using Re($U$) it is possible to identify if a cell undergoes a morphological change from a spherical to an elliptical shape. Thus, Re($U$) exhibits a significantly different profile along the optical axis for an elliptical object.

**3.3 Analyzing holograms of red blood cells**

To further test whether it was possible to determine axial changes by a complex biological specimen such as a red blood cell (RBC), which is biconcave, as well as to quantitate how Re(*U*) is changed by such object, we recorded physical holograms at distances from 30 to 100 µm (with fixed steps of 10 µm). RBCs were first immobilized to glass slides in a Phosphate Buffered Saline (PBS) solution keeping them in their natural shape, i.e., similar to a flat ellipse. The holograms were thereafter used for back-reconstruction and the positions of intensity maxima were evaluated and plotted versus the recording distance, i.e., the true position recorded by the translation of the piezo stage, which has sub-nm accuracy. The results of the experiments are presented as green dots in Fig. 4A. As can be seen from the data the position can accurately be determined by the reconstruction algorithm and the $R^2$ value of the linear fit was 0.99923. Furthermore, we also analyzed Re(*U*) of the light scattered by RBCs and a representative back-reconstructed profile along the optical axis from a hologram taken at 80 µm from the object is shown as a green solid line in Fig. 4B.

To compare the experiments of RBCs with simulated data a Cassini shaped model of a RBC with parameter values of, *a* = 2.2, *b* = 2.25, and *c* = 0.66 µm representing a 6.3 µm wide RBC-like object with an homogenous index of refraction of 1.40 was generated in Zemax [24]. The inset in Fig. 4A show a mesh plot of the Cassini model of the RBC used in the simulation. The geometry of the RBC object was described using a grid with triangular surfaces. This resulted in a not perfectly smooth surface, implying that noise was generated in the simulated diffraction pattern in form of thin fringes. When the synthetic hologram images from the simulation were back-reconstructed, oscillations of the intensity profile occurred close to the hologram plane (for reconstruction distances < 30 µm), both in intensity and phase. We therefore excluded the reconstructed data at short distances (<30 µm). However, this did not affect the outcome of the analysis, since the phase transition is located at much longer distances (~60µm) from the hologram. The result of the peak values of the intensity profiles from the back-reconstruction is shown as red dots in Fig. 4A. As can be seen, the maxima of the back-reconstructed intensity profiles, from the simulated holograms agree well with the corresponding experimental data, the $R^2$ value is 0.99679. In addition to the analysis of the intensity distribution we investigated the back-reconstructed profile of Re(*U*) for the RBC and simulated RBC, see the green and red curve in Fig. 4B, respectively.

As can be seen, the red and green curve representing the Re(*U*) agree well between 30 µm and 70 µm, particularly the dip at ~50 µm and the peak at ~60 µm. Comparing these results with that of the simulation of the ellipses, see Fig. 3B, indicates that the first peak is similar to the small ellipse indicating a weak focus far away from the cell. In Zemax this focus was found ~50 µm from the object, and the second indicates a strong focus close to the cell.

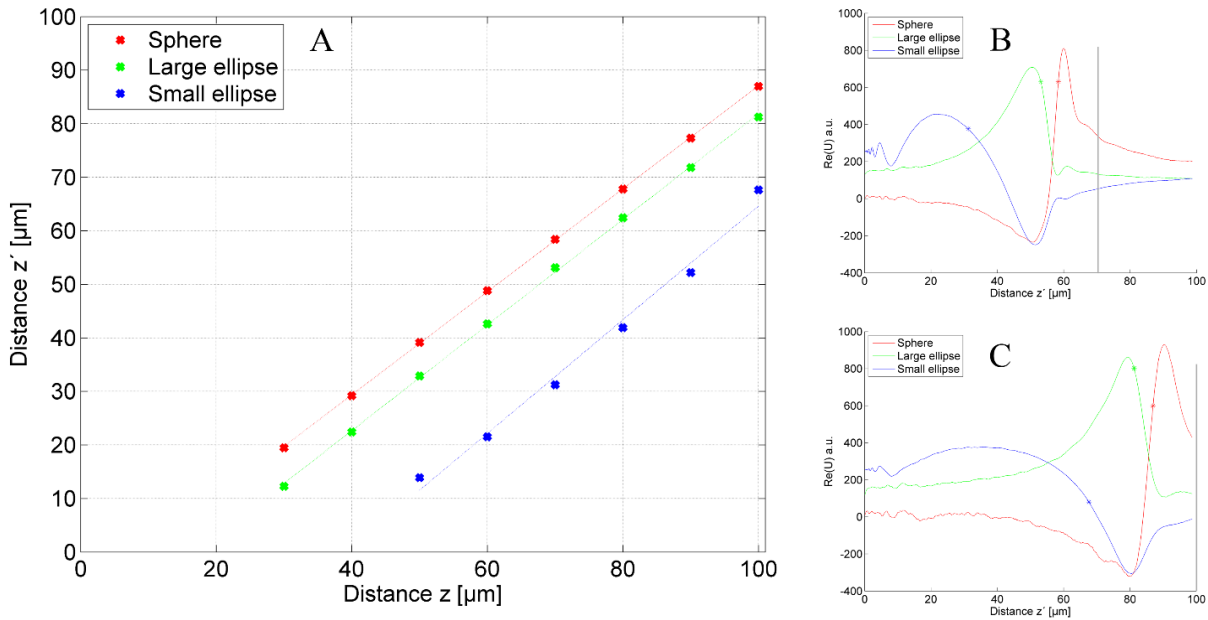

Figure 3. Simulation of elliptical shapes. Panel A) shows the reconstructed distance to the focus of the sphere (red), large ellipse (green), and the small ellipse (blue). Panel B) and C) show the Re(*U*) shift for the three objects at a distance of 70 µm and 100 µm respectively. The vertical grey lines represent the position of the object and the * indicates the location of the geometrical foci.

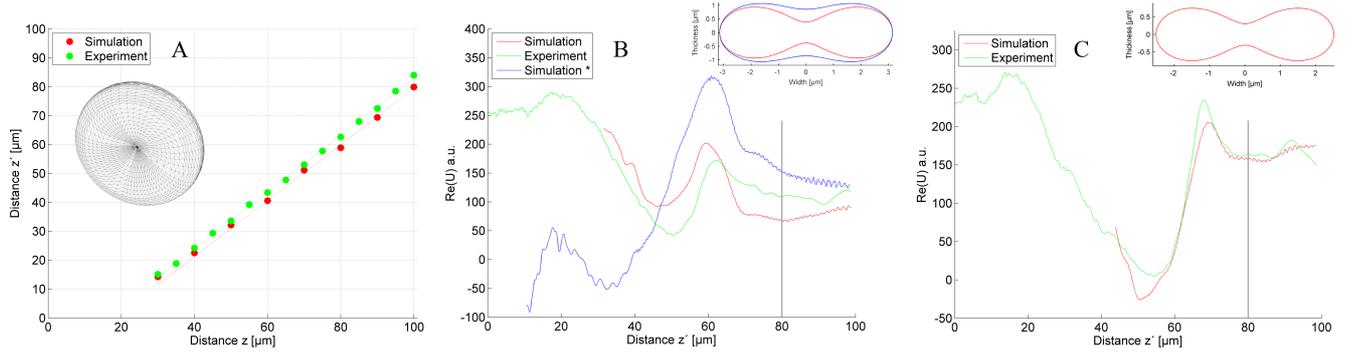

Figure 4. Panel A shows the reconstructed distance to the focus of the simulated RBC (red) and the experiment using a real RBC (green). The inset shows the Cassini model of the RBC used in the simulations. Panel B shows the reconstructed Re($U$) for the simulated unaltered RBC and a deformed RBC represented by the red and blue curves, respectively and for the experiment using a real RBC (green). Panel C shows reconstructed data from a ~20 % smaller RBC compared to the one used in panel B. Red and green curves represent the simulation and experiment, respectively. The grey vertical line in panel B and C indicates the position of the RBC.

We also investigated how a small morphological change of the RBC affected the Re($U$). The parameters of the Cassini model were changed to $a$ = 2.1, $b$ = 2.35, and $c$ = 0.66 μm making the RBC slightly deformed, see the inset in Fig. 4B for a cross-section profile. The peak values of the back-reconstructed intensity profiles for the deformed RBC overlapped well with that of the RBC (data not shown), implying that fluctuations of the position along the z-axis still could be accurately determined. However, the Re($U$) of the deformed RBC changed in comparison to the unaltered RBC, see the blue curve in Fig. 4B. In particular, the peak at ~60 μm is larger and the dip at 50 μm is shifted to ~35 μm as well as lowered.

To further investigate the robustness of the method we reconstructed data from a ~20 % (in comparison to the RBC used in Fig. 4B) smaller RBC and Cassini shaped model of a RBC, with the latter having parameter values of; $a$ = 1.76, $b$ = 1.80, and $c$ = 0.66 μm, representing a 5.04 μm wide object. The back-reconstructed profile of Re($U$) is presented in Fig. 4C and shows good agreement.

## 4. Discussion and Conclusion

The aim of this work was to present a simple, cost-effective and fast DHM method that can be used in flow chamber assays, and to investigate the possibility of using this to study rapid morphological changes of cells. Synthetic holograms and experimental data were used and the results from these were compared. Objects of various shapes; spherical, elliptical and RBCs-like were illuminated with a LED and the resulting holograms were recorded from 10 – 100 μm behind the objects. These were thereafter numerically back-reconstructed using the Rayleigh-Sommerfeld light propagation theory. From a stack of back-reconstructed images, using a z´-resolution of 70 nm, we first extracted the intensity profile at the x = 0 and y = 0 position in each image and created an intensity vs. distance plot along the z'-axis, as shown in Fig. 2B. The maximum values in each reconstructed stack of images were thereafter plotted versus the actual sampled hologram position, giving rise to plots as shown in Fig. 2E. For all analyzed objects, it was found that the reconstruction was very accurate providing nice linear fits to the data.

Since it is not possible to use the intensity profile nor the pure phase shift of the field to quantitate a morphological change of a cell we evaluated the real part of the complex amplitude Re($U$). In Re($U$) the Gouy phase shift is taken into account providing more information of the change of the field caused by an object. The Gouy phase shift is the result of an additional phase contribution $\phi_G$ to the propagating phase $k \cdot n \cdot z$ ($k$ being the wavevector, $n$ the refractive index and $z$ the distance) a light wave experiences as it travels through its geometrical focus. Back-reconstruction of the holograms with evaluation of Re($U$) along the optical axis thus give information of the shape of an object, or can be used to analyze geometrical changes an object undergoes during a certain time period (e.g. a biological cell that collapses over time).

We first compared Re($U$) from simulated objects of different degrees of ellipticity and concluded that Re($U$) significantly differed between these. Moreover, we ran a set of experiments with spherical objects and RBCs and in turn, compared these data with simulated. The maxima of the back-reconstructed profiles for the simulated and the experimental RBC agree very well, except for oscillations in the simulated RBC (close to the hologram plane). When comparing the shape of the phase shift of the RBC (Fig. 4B) with the phase shifts for the sphere and ellipses (Fig. 3B), they showed some similarities. From 0 to ~50 mm, the shape is similar to the medium ellipse, and after ~50 μm, the shape appears more like the shape of the sphere. This suggests that the RBC creates two geometrical focuses that appear as a superposition in the back-reconstructed profile. This also agrees with the simulated results obtained directly from Zemax. This means that the shape of the phase shift can be regarded as a "signature" of how strongly an object is focusing the incoming light.

More noise was observed in the back-reconstructions from the simulated RBC-like object in comparison to the experimental. This originates from the fact that the surface of the simulated object is geometrically assembled by a number of triangles, and is thus not a perfect continuous surface. Even though the shape of the model is not perfect, the results using the Cassini model fits well the experimental results of the RBC well for distances >30 μm. In addition, we changed the parameter values of the Cassini model and made a RBC with a deeper center part. This did not significantly change the way the light was focused, however, the Re($U$) was significantly changed, indicating that by only observing the Re($U$) it is possible to identify a morphological change. Also, good agreement between experimental and simulated data was found when reconstructing data of a ~20 % smaller RBC, which thus indicates that the method is robust.

Finally, the simple setup and fast way of reconstructing the data presented in this work has some restrictions that the reader needs to be aware of before implementation. The quality of the hologram depends on the difference in refractive index between the object and the surrounding media. For example, the reconstruction of a Gabor hologram also produce a conjugate image overlaid the real image. This is, however, completely out of focus and has a significantly lower amplitude than the reconstructed field but needs to be considered for very weak scattering objects, such as when imaging objects with refractive indices close to the surrounding media. Having control of the orientation of the object throughout the measurement is also important, since rotation of a non-homogenous objects can change Re($U$).

In conclusion, in this work we investigated the possibility of using DHM for identification of the geometrical shape of a cell using a cheap LED and a simple setup around an inverted microscope. We found that Re($U$) provided sufficient information to distinguish dissimilar geometrical objects, which was confirmed using both simulated and experimental images. The position of these objects could also accurately be determine along the optical axial. Thus, this method can be used to measure real-time morphological changes of living cells, in e.g., optical tweezers cell experiments or flow chamber assays, for pathophysiological studies when fast transitions are monitored in real-time at different heights from a surface.

## 5. Acknowledgement

This work was supported by the Swedish Research Council (621-2013-5379) to M.A. We are grateful to Bhupender Singh for help with preparation of RBCs and Prof. Ove Axner for valuable discussions and comments.